\documentclass[conference]{IEEEtran}
\usepackage[utf8]{inputenc}
\usepackage{amsbsy}
\usepackage{epsfig}
\usepackage{float}
\usepackage{graphicx}
\usepackage{color}
\usepackage{url}
\usepackage{subcaption}
\IEEEoverridecommandlockouts

\usepackage{cite}
\usepackage{amsmath,amssymb,amsfonts}
\usepackage[caption=false]{subfig}
\usepackage[justification=centering]{caption}
\usepackage{algorithmic}
\usepackage[ruled,vlined]{algorithm2e}
\usepackage{textcomp}
\usepackage{xcolor}
\usepackage{balance}
\usepackage{booktabs}
\usepackage{textcomp, flushend}

\newtheorem{remark}{Remark}

\title{Toward Integrated Sensing, Communications,\\ and Edge Intelligence Networks\thanks{Corresponding author e-mail: mattia.merluzzi@cea.fr. This work has been partly supported by the SNS JU project 6G-GOALS under the EU’s Horizon program Grant Agreement No 101139232. Figure~\ref{fig:sys_model} was partly designed using resources from Flaticon.com. Also, although designed by the authors, the left part was generated with the assistance of ChatGPT (OpenAI).}\vspace{-.2cm}}
\author{Mattia Merluzzi$^1$, Miltiadis C. Filippou$^2$, Paolo Di Lorenzo$^{3,4}$, and George C. Alexandropoulos$^5$\\
$^1$CEA-Leti, Université Grenoble Alpes, F-38000 Grenoble, France\\
$^2$WINGS ICT Solutions, 17121 Athens, Greece\\
$^3$CNIT, 43124 Parma, Italy; $^4$Sapienza University of Rome, 00184 Rome, Italy\\
$^5$National and Kapodistrian University of Athens, 16122 Athens, Greece \vspace{-0.3 cm}}

\begin{document}
\maketitle

\begin{abstract}
Wireless systems are expanding their purposes, from merely connecting humans and things to connecting intelligence and opportunistically sensing of the environment through radio-frequency signals. In this paper, we introduce the concept of triple-functional networks in which the same infrastructure and resources are shared for integrated sensing, communications, and (edge) Artificial Intelligence (AI) inference. This concept opens up several opportunities, such as devising non-orthogonal resource deployment and power consumption to concurrently update multiple services, but also challenges related to resource management and signaling cross-talk, among others. The core idea of this work is that computation-related aspects, including computing resources and AI models availability, should be explicitly considered when taking resource allocation decisions, to address the conflicting goals of the services coexistence. After showing the natural coupling between theoretical performance bounds of the three services, we formulate a service coexistence optimization problem that is solved optimally, and showcase the advantages against a disjoint allocation strategy.
\end{abstract}

\begin{IEEEkeywords}
Goal-oriented semantic communications, adaptive computation, resource allocation, spectrum sharing, ISAC. 
\end{IEEEkeywords}

\section{Introduction}
Two technologies are recognized today as key enablers of the upcoming sixth Generation (6G) of wireless networks: (i) Integrated Sensing And Communications (ISAC) \cite{6G-DISAC-magazine}, and (ii) Semantic and goal-oriented Communication (SemCom) \cite{6G_GOALS,di2023goal}. The former is related to opportunistically using communication radio-frequency signals to sense the environment (e.g., radar-based target parameter estimation), while the latter focuses on transmitting only the relevant information for the purpose of a task. This feature is especially relevant when connecting Artificial Intelligence (AI) agents that run inference on capillary data for, e.g., monitoring complex environments. 

ISAC is enabled by flexible duplexing capabilities at the Base Stations (BSs) and novel waveforms among others \cite{6G-DISAC-magazine,10769781}. The major design challenge for this technology is how to split wireless resources between communications and sensing (e.g., power allocation and beamforming design \cite{Fan22}). On the other hand, SemCom has been made possible by the recent advances in the field of AI, and the deployment of distributed computing resources within the networks, thereby, enabling speedy processing of local, information-rich data. However, the main challenge for SemCom is the cost split between communications and computation. For instance, per one approach, the data source may aggressively compress data to transfer to an Edge Server (ES), thus saving wireless resources, while, on the other side, requiring that the ES makes use of a model robust to different data compression ratios to output the inference result at the cost of increased computing resource consumption~\cite{Merluzzi2025}. This naturally generates a cross-layer decision space that covers all layers of the communication protocol stack from the physical up to the application layer, and needs to account for both user device and network infrastructure capabilities.
When coupling ISAC and SemCom, an even larger decision space arises, including resource allocation, multiple access schemes, AI model selection, and data representation schemes.\\ 
\textbf{Related works.} Resource sharing is a natural key feature of ISAC, as the goal is to opportunistically use radio transmission for sensing \cite{6G-DISAC-magazine}. 
In \cite{Fan22}, the Cram\'{e}r-Rao bound (CRB) for angle estimation is derived as a function of the beamforming vector, which is then optimized with the goal of minimizing the CRB under a data rate constraint for the served user. On another recent ISAC example, \cite{Simultaneuous_CaS} presented hybrid analog and digital beamforming designs for Full-Duplex (FD) Multiple-Input Multiple-Output (MIMO) systems that optimize downlink communications while guaranteeing a maximum target position error bound threshold.
\begin{figure*}[t]
    \centering
    \includegraphics[width=0.85\textwidth]{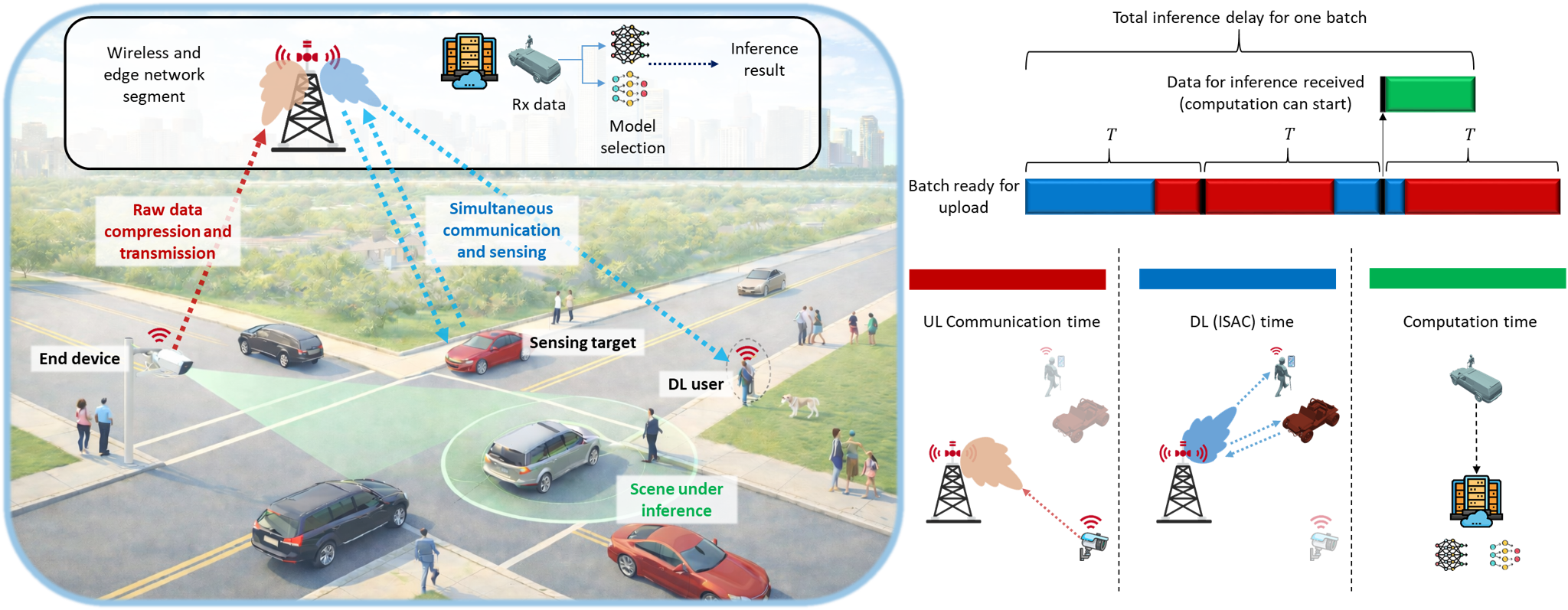}
    \caption{The considered multi-service wireless system and frame structure.}
    \label{fig:sys_model}
\end{figure*}
On the other hand, SemCom enables new ways of spectrum sharing~\cite{Duong2011CooperativeSpectrumSharing}, mainly because the capability of achieving a goal based on exchanged data (goal effectiveness) depends on their quality (affected by compression loss along with data perturbations owing to co-channel interference and reception noise), along with the availability of computing resources. A dynamic spectrum sharing mechanism between a goal-oriented and a legacy user was proposed in~\cite{Merluzzi2025}, where availability of more or less powerful (and thus resource consuming) AI models  varied over time due to average computational load constraints. Other works focus on non-orthogonal multiple access schemes to improve the achievable rate for a legacy user, thanks to radio resource sharing with a SemCom user \cite{Xidong23}. These works focus on the coexistence between SemCom and legacy users mainly seeking to maximize their throughput. 
Finally, recent works have started exploring the presence of the three services \cite{Kailin2026,Dingzhu25, Dingzhu24}, but solely focusing on Edge Inference (EI) based on sensing data, and not on resource sharing for service coexistence.

\noindent\textbf{Our contribution.} With this paper, we go one step further investigating the problem of how network resources should be shared among sensing, communications, and EI services/tasks. In particular, we propose a computation-aware resource sharing scheme improving the trade-off between ISAC costs and EI performance. Given system information in terms of in-network AI model availability (relevant to EI) and their performance, our objective is to jointly optimize the transmit power for ISAC and the EI input data representation format to realize effective service coexistence. Toward this end, we formulate a mixed integer non-linear problem that is solved \textit{optimally}.
\section{System Model and Services Metrics}\label{sec:sys_model}
We consider the wireless scenario illustrated in Fig.~\ref{fig:sys_model}, in which a BS provides resources to UpLink (UL) communications for an EI service involving a camera uploading images for classification or object detection. At the same time, other resources are offered for simultaneous DownLink (DL) communications and monostatic-type sensing~\cite{Simultaneuous_CaS}. In particular, there exists a single-antenna user (DL user) requesting, e.g., video streaming and a point-source target in the vicinity of the requesting user. For this purpose, the BS deploys an FD MIMO architecture~\cite{9933358} with $N_t$ and $N_r$ transmit and receive antenna elements, respectively. 
Time is assumed to be organized in frames $t=1,2,\ldots$ of equal duration $T$. During each $t$-th frame, a portion $\rho_{\rm ul}[t]\in[0,1]$ is used for UL communications, as needed for the camera to upload inference input data, while the remaining portion $\rho_{\rm dl}[t]\in[0,1]$ is dedicated to ISAC. For each $t$-th frame, it is assumed that $\rho_{\rm ul}[t]+\rho_{\rm dl}[t]=1$. 

As shown in Fig.~\ref{fig:sys_model}, in the general case, the total delay to serve one uploaded data batch, from the end of batch data generation\footnote{We assume that an on-camera data batch is scheduled for uploading, starting from the frame that initiates immediately after the end of batch generation, either in the beginning or after the DL subframe, depending on traffic type prioritization.} till the time inference output is issued, incorporates three non-contiguous time interval types, namely: \textit{i}) the total buffering time for the end device, owing to prioritized DL transmissions to the streaming user; \textit{ii}) the time needed to upload the batch to the ES, the end of which is marked by an ``end of input'' symbol at the end of the respective frame; and \textit{iii}) the batch inference (processing) time.

The communication protocol is, in fact, ``quasi-FD,'' in the sense that FD capability of the BS only applies to signaling time relating to the ISAC service (i.e., during DL sub-frames), whereas the BS operates in reception-only (half-duplex) mode during data upload by the end device (camera). During time slots dedicated to  UL, data (either raw video frames or compressed versions thereof) are uploaded by the camera to the BS collocated with an ES hosting a Machine Learning (ML) model that performs an inference task, such as object (e.g., vehicle) classification. Specifically, an ML model hosted by the ES is assumed to operate on data batches of size $B$ samples. Instead, during time slots devoted to ISAC service signaling, the DL user is served, while echoes of the DL data signals bouncing from a target (object) in the vicinity of the user are, almost concurrently, received at the BS and used for its parameters estimation. All services are carried out under a multi-carrier setup with $F$ SubCarriers (SCs), with $W$ being the SC spacing.

\begin{remark}
One may wonder why, since the BS has FD capability, such capability is only exploited partially, namely, to enable simultaneous DL communications (BS's transmit mode) and sensing (BS's receive mode) during ISAC sub-frames, but not to combine ISAC and UL signaling within the same time resources~\cite{FD_MIMO_ISAC_ULDL}. The reason is that UL transmissions carrying camera data are typically several tens of dB stronger than the weak radar echoes of interest, and would, therefore, dominate the receiver’s dynamic range, distort the echo covariance structure, and severely degrade sensing accuracy, even under aggressive self‑interference cancellation and spatial nulling~\cite{9933358}. For this reason, UL signaling is orthogonalized in time, whereas FD operation is reserved for the ISAC functionality, where the BS must only suppress its own transmitted waveform to recover environmental echoes.
\end{remark}

In the following, we present the Key Performance Indicators (KPIs) for the EI service, initiated by the UL raw data communications during the respective frame portion, as well as the ISAC metrics governing the frame portion for simultaneous DL data communications and monostatic-type sensing.

\subsection{Inference Key Performance Indicators}
\subsubsection{Inference Delay}
The inference delay depends on data compression and throughput. Let $\mathbf{w}_{{\rm ul},f}[t]\in\mathbb{C}^{N_r\times 1}$ denote the combining vector and $\mathbf{h}_{{\rm ul},f}[t]\in\mathbb{C}^{N_r\times 1}$ the UL channel response vector between the device and the BS, both during the $t$-th time frame and on SC $f$. Then, the Signal-to-Noise Ratio (SNR) on each $f$-th SC during time frame $t$ reads as:
\begin{equation}\label{SNR}
\text{SNR}_{{\rm ul},f}[t]\triangleq\frac{P_{\text{ul},f}[t]\left|\mathbf{w}_{{\rm ul},f}^{\rm H}[t]\mathbf{h}_{{\rm ul},f}[t]\right|^2}{N_0W},
\end{equation}
where $P_{\text{ul},f}[t]$ represents the UL transmit power in each $f$-th SC during each $t$-th frame, and $N_0$ denotes the noise power spectral density. Hence, the average effective UL data rate during time frame $t$ can be expressed in bits/sec as follows:
\begin{equation}\label{rate_ul}
    R_{\rm ul}[t]\triangleq\rho_{\rm ul}[t]W\sum\nolimits_{f=1}^F \log_2\left(1+\text{SNR}_{{\rm ul},f}[t]\right).
\end{equation}

Let $n_b$ represent the number of bits in each $b$-th UL data batch ($b=1,2,\ldots,B$); recall that inference takes place at the ES when a full batch is received. Considering a persistent Time Division Duplexing (TDD) of UL and DL, we approximate the delay in seconds to upload each batch at the ES as follows:
\begin{equation}\label{comm_delay}
    L_{\text{comm},b}\triangleq\left\lceil\frac{n_b}{\overline{R}_{{\rm ul},b}T}\right\rceil T\approx \frac{n_b}{\bar{R}_{{\rm ul},b}},
\end{equation}
where $\bar{R}_{{\rm ul},b}$ is the average data rate across frames (i.e., averaging values, each of which is given by \eqref{rate_ul}) during each $b$-th batch upload. Finally, denoting by $L_{\text{comp},b,m}$ the computation delay to infer one batch using a ML model $m$, the total per-batch delay from the initiation of data collection and till the inference output is issued is:
\begin{equation}
    L_{\text{tot},b}\triangleq L_{\text{comm},b}+L_{\text{comp},b,m}.
\end{equation}

\subsubsection{Goal Effectiveness}
The inference delay is insufficient to provide a complete assessment of the inference performance. To this end, let $Q_b(n_b,m)$ indicate a generic inference quality metric for batch $b$, which is a (typically monotonic) function of the data quality represented by the number of bits $n_b$ and the ML model $m$ used for inference. For example, $Q_b(n_b,m)$ can denote the number of correctly classified samples in a batch, or the inference confidence. Let also $Q_{\min}$ and $L_{\max}$ denote, respectively, a minimum quality level and a maximum tolerable delay. By assuming that an inference task per $b$-th batch is successful if the minimum quality is achieved (i.e., when $Q_b(n_b,m)\geq Q_{\min}$) within a delay $L_{\text{tot},b}\leq L_{\max}$, we define the goal effectiveness for a specific ML model $m$ as: 
\begin{equation}\label{go_eff}
    E_g\triangleq\mathbb{E}\left\{\mathbf{1}\left(Q_{b}(n_b,m)\geq Q_{\min}\right)\times\mathbf{1}\left(L_{\text{tot},b}\leq L_{\max}\right)\right\},
\end{equation}  
where $\mathbf{1}(\cdot)$ represents the indicator function, while, the expectation in the left-hand side is taken with respect to random context parameters, such as wireless channels, inference input data batches, inference delay, and positions of users.

\subsection{ISAC Metrics}
\subsubsection{Achievable Rate}
Let $\mathbf{w}_{{\rm dl},f}[t]\in\mathbb{C}^{N_t\times 1}$ denote the precoding vector and $\mathbf{h}_{{\rm dl},f}[t]\in\mathbb{C}^{1\times N_t}$ the DL channel response vector between the BS and the DL user, both on each $f$-th SC during each $t$-th time frame. Consequently, for each of those frames, the SNR on each $f$-th SC and the average effective DL rate in bits/sec are given, respectively, by:
\begin{align}\label{rate_dl}
    &\text{SNR}_{{\rm dl},f}[t]\triangleq\frac{P_{\text{dl},f}[t]\left|\mathbf{h}_{{\rm dl},f}[t]\mathbf{w}_{{\rm dl},f}[t]\right|^2}{N_0W},\nonumber\\
    &R_{\rm dl}[t] \triangleq \rho_{\rm dl}[t]W\sum\nolimits_{f=1}^F\log_2\left(1+\text{SNR}_{{\rm dl},f}[t]\right),
\end{align}
where $P_{\text{dl},f}[t]$ represents the DL transmit power on SC $f$.

\subsubsection{Cramér-Rao Bound of Target Estimation} Let $\mathbf{h}_{f,\mathrm{tx}}[t,\theta]\in\mathbb{C}^{1\times N_t}$ and $\mathbf{h}_{f,\mathrm{rx}}[t,\theta]\in\mathbb{C}^{N_r\times 1}$ represent the steering vectors between the BS transmit antennas and the target as well as the target and the BS receive antennas, respectively, both depending on the angle of departure/arrival $\theta$ (the transmit and receive antenna arrays at the BS are considered sufficiently spaced at known distance~\cite{9933358}, enabling to assume a common angle of departure and arrival). The target response on each $f$-th SC during each $t$-th frame is modeled by the following rank-one MIMO channel matrix:
\begin{equation}
\mathbf{H}_f[t,\theta]
=
\alpha_{f}[t]\;\mathbf{h}_{f,\mathrm{rx}}[t,\theta]\mathbf{h}_{f,\mathrm{tx}}[t,\theta],
\label{eq:H_rank1}
\end{equation}
where $\alpha_{f}[t]$ contains the unknown complex-valued round-trip attenuation that depends on the target's Radar Cross Section (RCS). Note that reflections from the user have been ignored in~\eqref{eq:H_rank1} as either being highly attenuated or extracted from the overall echo channel via a dedicated user positioning process~\cite{Simultaneuous_CaS}. Let, finally, vector \(\boldsymbol{\eta}_f[t] \triangleq [\theta,\alpha_{{\rm R},f}[t],\alpha_{{\rm I},f}[t]]^{\rm T}\),
where \(\alpha_f[t]\triangleq\alpha_{{\rm R},f}[t]+\jmath\alpha_{{\rm I},f}[t]\), include the unknown target-induced parameters per $k$-th SC during each $t$-th time frame.

Let us assume that, at each \(f\)-th SC during each \(t\)-th time frame, the BS transmits the $K[\rho_{\rm dl}[t]]$ data symbols \(\mathbf{s}_f\in\mathbb{C}^{1\times K[\rho_{\rm dl}[t]]}\) in the DL direction, which are normalized as \(K^{-1}\mathbf{s}_f\mathbf{s}_f^{\rm H}=1\). Note that, in general, the number of DL data symbols depends on the portion of time $\rho_{\rm dl}[t]$ devoted to DL transmissions. Considering adequate FD MIMO operation~\cite{10769781}, the reflections of these symbols from the target are received back at the \(N_r\) BS receive antennas as $\mathbf{Y}_f[t,\theta]
\triangleq
\mathbf{H}_f[t,\theta]\;\mathbf{w}_{{\rm dl},f}[t]\; \mathbf{s}_f
+
\mathbf{N}_f[k],$
where \(\mathbf{N}_f\sim\mathcal{CN}(\mathbf{0}_{N_r\times K[\rho_{\rm dl}[t]]},N_0 W\mathbf{I}_{N_r})\) denotes
circularly symmetric complex Gaussian noise, which is assumed independent across
antennas, snapshots, and SCs. Clearly, using \(\mathbf{M}_f[t,\theta]\triangleq
\mathbf{H}_f[t,\theta]\mathbf{w}_{{\rm dl},f}[t]\mathbf{s}_f\), the $N_r\times K[\rho_{\rm dl}[t]]$ matrix $\mathbf{Y}_f[t,\theta]$ is distributed as \(\mathcal{CN}(\mathbf{M}_f[t,\theta],N_0 W\mathbf{I}_{N_r})\).

Following~\cite{kay1993fundamentals}, the elements of the $3\times3$ Fisher Information Matrix
(FIM) for \(\boldsymbol{\eta}_f[t]\), $\mathbf{J}_{f}[t]$, are defined as ($i,j=1,2,3$):
\begin{align*}
[\mathbf{J}_{f}[t]]_{i,j}\triangleq\frac{2}{N_0 W}
\Re\left\{{\rm Tr}\left\{
\left(\frac{\partial \mathbf{M}_f^{\rm H}[t,\theta]}{\partial [\boldsymbol{\eta}_f[t]]_i}\right)
\left(\frac{\partial \mathbf{M}_f[t,\theta]}{\partial [\boldsymbol{\eta}_f[t]]_j}\right)
\right\}\right\}.
\end{align*}
By treating $\alpha_{{\rm R},f}[t]$ and $\alpha_{{\rm I},f}[t]$ as nuisance parameters and using the $N_r$-element vector definitions $\mathbf g_{f}[t,\theta] \triangleq \mathbf h_{f,\mathrm{rx}}[t,\theta]\mathbf h_{f,\mathrm{tx}}[t,\theta]\mathbf w_{{\rm dl},f}[t]\sqrt{P_{\text{dl}, f}[t]}$ and $\dot{\mathbf g}_{f}[t,\theta] \triangleq \frac{\partial \mathbf g_{f}[t,\theta]}{\partial \theta}$, the equivalent FIM for the unknown target parameter $\theta$ can be calculated via the Schur complement, as follows:
\begin{align}
\bar{J}_{f,t}(\theta)
\triangleq
\frac{2K[\rho_{\rm dl}[t]]}{N_0 W}\left|\alpha_f[t]\right|^2
\dot{\mathbf g}_{f}^{\rm H}[t,\theta]
\mathbf P^\perp_{\mathbf{g}}
\dot{\mathbf g}_{f}[t,\theta],
\end{align}
where
\(\mathbf P^\perp_{\mathbf{g}}
\triangleq
\mathbf{I}_{N_r\times N_r}-\frac{\mathbf g_{f}[t,\theta]\mathbf g_{f}^{\rm H}[t,\theta]}
{\|\mathbf g_{f}[t,\theta]\|^2}\). Finally, aggregating information across all \(F\) SCs per $t$-th time frame, the CRB for the estimation of the target parameter $\theta$ is computed as:
\begin{align}\label{crb}
\mathrm{CRB}_\theta[t] 
\triangleq \left(\sum\nolimits_{f=1}^{F}\bar{J}_{f,t}(\theta)\right)^{-1}=\frac{N_0 W}{2K[\rho_{\rm dl}[t]]\gamma},
\end{align}
where we have used the definition:
\begin{align}
\gamma\triangleq\sum\nolimits_{f=1}^{F}\left|\alpha_f[t]\right|^2\nonumber
\dot{\mathbf g}_{f}^{\rm H}[t,\theta]\mathbf P^\perp_{\mathbf{g}} \dot{\mathbf g}_{f}[t,\theta].
\end{align}
Note that, for the case of pulses of duration $W^{-1}$, $K[\rho_{\text{dl}}[t]]=\rho_{\text{dl}}[t]TW$ holds.

\section{The Coexistence of the Three Services}
In this section, we commence by elaborating on the interrelation between the considered ISAC and EI computing aspects. Let us assume that, for batch $b$, a ML model $m$ is used for inference at the ES. We impose the following condition to meet the latency constraint $L_{\max}$:
\begin{equation}
\rho_{\text{ul}}[t]\geq\displaystyle \frac{n_b}{L_{\text{comm}}^*W\sum_{f=1}^F \log_2\left(1+\text{SNR}_{{\rm ul},f}[t]\right)}, 
\end{equation}
where $L_{\text{comm}}^* \triangleq L_{\max} - L_{\text{comp},b,m}$ denotes the remaining latency budget after accounting for computation. This immediately yields the following constraint on the DL time allocation:
\begin{equation}\label{rho_dl}
    \rho_{\text{dl}}^*[t]=1-\frac{n_b}{L_{\text{comm}}^*W\sum_{f=1}^F \log_2\left(1+\text{SNR}_{{\rm ul},f}[t]\right)}.
\end{equation}
Recalling \eqref{rate_dl} and \eqref{crb}, the CRB for the target parameter estimation can be explicitly expressed as a function of the communication and computation resources associated with the inference service (cf. \eqref{rho_dl}), as follows:
\begin{align}\label{crb_f_of_ei}
        \displaystyle \mathrm{CRB}_\theta[t] 
        =\frac{N_0}{2T\rho_{\text{dl}}^*[t] \gamma}.
\end{align}
In the case of angle estimation for a point target, we have \cite{Fan22}:
\begin{equation}
\gamma=\sum\nolimits_{f=1}^{F}\left|\alpha_f[t]\right|^2 ||\dot{\mathbf h}_{\text{rx}}[t,\theta]||^2|\mathbf{h}_{\text{tx}}[t]\mathbf{w}_{\text{dl}}[t]|^2P_{\text{dl},f}[t].
\end{equation}
Similarly, we can write the DL user data rate as follows:
\begin{equation}\label{rate_dl_f_of_ei}
        \displaystyle R_{\rm dl}[t] = \rho_{\text{dl}}^*[t] W\sum_{f=1}^F\log_2\left(1+\text{SNR}_{\text{dl},f}[t]\right).
\end{equation}

Putting all above together, expressions \eqref{rho_dl}, \eqref{crb_f_of_ei}, and \eqref{rate_dl_f_of_ei} jointly reveal the fundamental three-way coupling among communication, computation, and sensing functionalities: the selected inference model determines the available communication time through $L_{\text{comp},b,m}$, which in turn fixes the ISAC time allocation $\rho_{\text{dl}}[t]$, thereby simultaneously shaping both the sensing accuracy and the achievable DL data rate.

\subsection{Problem Formulation}
We now formulate a design optimization problem for the considered multi-service wireless system, aiming to explore the trade-off between ISAC and EI performance, with \textit{compute-awareness} being a key feature of the study. This means that the specific ML model used for inference affects this trade-off, and should be taken into account when allocating resources. The problem is mathematically formulated as follows with $\sigma\in[0,1]$ being a weighting parameter:
\begin{align*}
    \mathcal{P}: &\underset{\{P_{\text{dl},f}\}_{f=1}^F,n_b}{\min}\quad\sigma \frac{\sum_{f=1}^F P_{\text{dl},f}}{P_{\text{dl}}^{\max}}+(1-\sigma)\frac{n_b^{\max}}{n_b}\\
    &\hspace{0.545cm}\text{subject to}\,\,(a)\!\!: \text{CRB}_{\theta}\leq \text{CRB}_{\theta}^{\text{th}},\,\,(b)\!\!: R_{\text{dl}}\geq R_{\text{dl}}^{\text{th}},\nonumber\\
    &\hspace{2.08cm}(c)\!\!: P_{\text{dl},f}\geq 0\,\;\forall f,\,\,
    (d)\!\!: n_b\in\mathcal{N}_b,\\
    &\hspace{2.08cm}(e)\!\!: \sum\nolimits_{f=1}^F P_{\text{dl},f}\leq P_{\text{dl}}^{\max}.
\end{align*}
The first term in $\mathcal{P}$'s objective function aims at minimizing the BS transmit power allocated to the ISAC service, while the second term intends to maximize the accuracy of the EI service (i.e., richness of offloaded data representation). For the constraints: $(a)$ indicates that the CRB needs to be lower than a threshold $\text{CRB}_{\theta}^{\text{th}}$ (cf. \eqref{crb}, \eqref{crb_f_of_ei}); $(b)$ imposes a minimum data rate guarantee for the communication service (cf. \eqref{rate_dl_f_of_ei}); $(c)$ ensures that all DL transmit powers allocated on the SCs are non-negative; $(d)$ implies that the data representation belongs to the set $\mathcal{N}_b$; and $(e)$ means that BS operates in the DL with a transmit power upper bounded by $P_{\text{dl}}^{\max}$. Finally, the inference delay constraint is implicitly taken into account through \eqref{rho_dl}.

The problem $\mathcal{P}$ is a mixed integer non-linear program. However, once $n_b$ is fixed, it is a convex problem, due to the fact that its objective function is linear, and constraints $(a)$ and $(b)$ are convex; all other constraints are linear. Noting that $\mathcal{N}_b$ is a set with low cardinality, we can solve ${\rm card}(\mathcal{N}_d)$ convex problems, and select the optimal solution minimizing $\mathcal{P}$'s objective function. To this end, for each $n_b\in\mathcal{N}_b$, the convex problem can be solved with the CVXPY\footnote{https://www.cvxpy.org/index.html} Python-embedded modeling language.

\begin{figure*}[t]
    \centering
    \begin{subfigure}{0.44\textwidth}
        \centering
        \includegraphics[width=\linewidth]{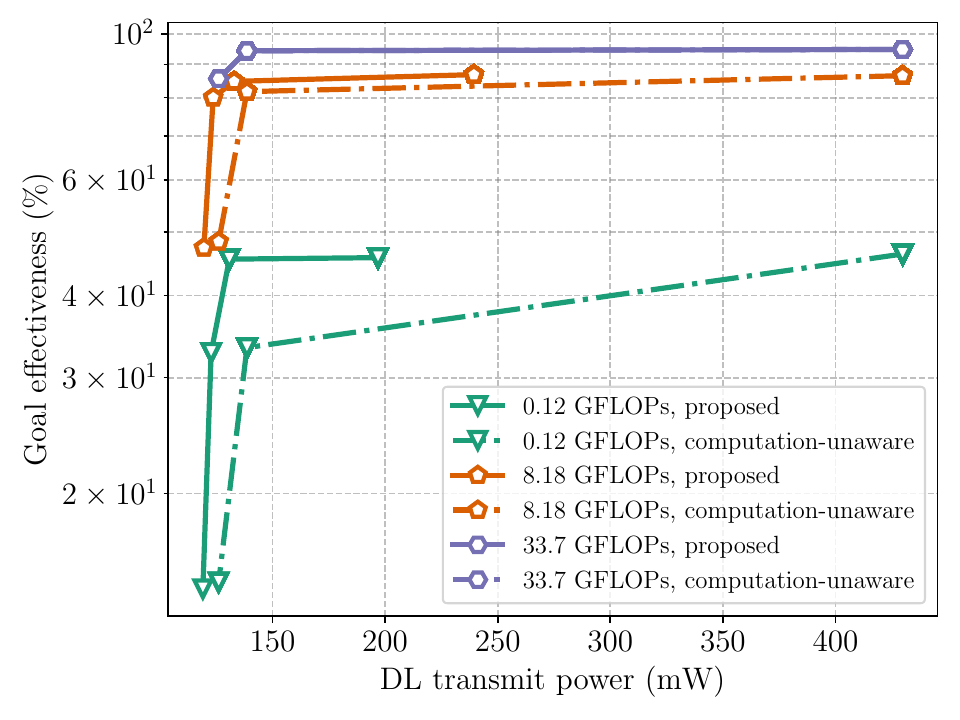}
        \caption{}
        \label{fig:tradeoff_1}
    \end{subfigure}
    \hspace{.5cm}
    \begin{subfigure}{0.44\textwidth}
        \centering
        \includegraphics[width=\linewidth]{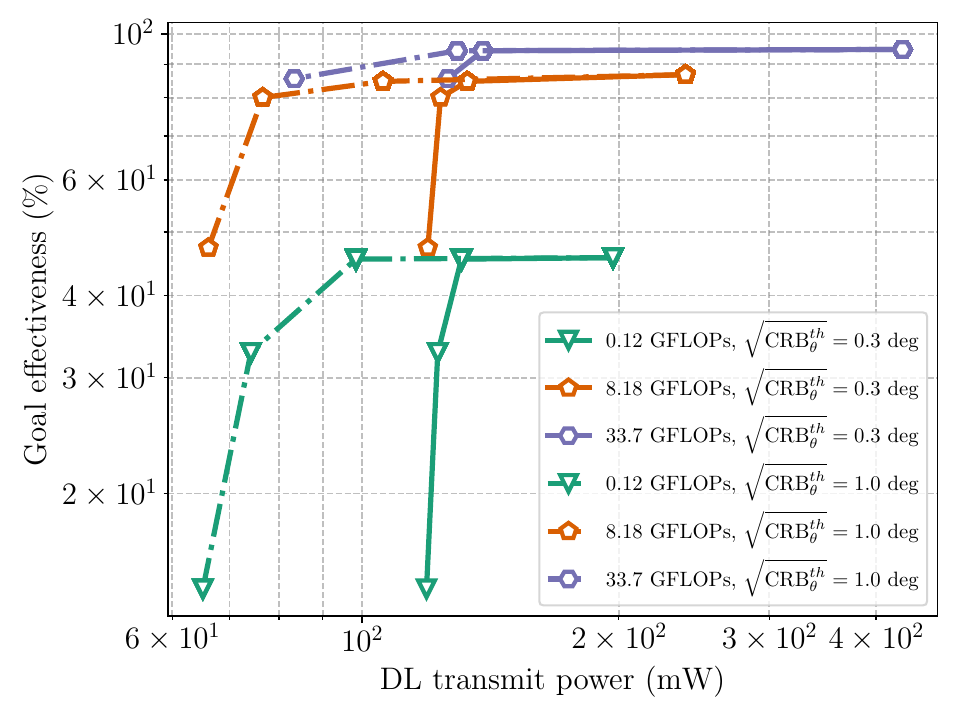}
        \caption{}
        \label{fig:tradeoff_2}
    \end{subfigure}
\caption{Goal effectiveness versus (vs.) DL transmit power: 
(a) Trade-off vs. $\sigma$ in $\mathcal{P}$; 
(b) Trade-off under two CRB constraints.\vspace{-.3cm}}
    \label{fig:tradeoff}
\end{figure*}

\section{Numerical Results and Discussion}
In this section, we assess the performance of the proposed multi-service wireless system design, starting with the description of the setting of parameters for all three services, followed by the requirements definition and the simulated benchmarks.

\textbf{Communication-related parameters.} We have considered transmission at carrier frequency $10$ GHz, SC spacing $W=30$ KHz, and a total bandwidth of $50$ MHz, i.e., $F=1666$ SCs. In the designed 2D scenario, the BS was located at $[0,0]$, the edge device at $[0,80]$, and the DL user (communication service) at $[50,55]$. The BS was equipped with a uniform linear array of $N_t=N_r=16$ antennas, spaced by $\lambda/2$ ($\lambda$ being the wavelength) along the azimuth. At all receivers, $N_0=-174$~dBm/Hz and a $10$~dB noise figure were considered. All channels were generated through the steering vectors as in \cite{Fan22}, with path loss $\left(\frac{\lambda}{4\pi}\right)^2d_{\text{tx},\text{rx}}^{-\beta},$ where $d_{i,j}$ is the distance between transmitter and receiver, and $\beta=2.5$ the path loss exponent. The total transmit power of the UL device was set as $P_{\text{ul}}=0.1$~W, equally allocated across all SCs. Maximum ratio transmission ($\mathbf{w}_{\text{dl}}[t]=\mathbf{h}_{\text{dl},f}^{\rm H}/||\mathbf{h}_{\text{dl},f}||$) and combining ($\mathbf{w}_{\text{ul}}[t]=\mathbf{h}_{\text{ul},f}/||\mathbf{h}_{\text{ul},f}||$) were are adopted at the FD BS. 

\textbf{Sensing-related parameters.} The sensing target was located at $[20,20]$ and the RCS was set to $1$ m$^2$.

\textbf{Inference-related parameters.} The inference task was image classification performed on CIFAR-100 \cite{krizhevsky2009}. The images were resized to $224\times 224$ pixels (i.e., $3$ channels). At the device, an autoencoder with four convolutional layers (each downsampling by $2$) plus an adaptive bottleneck layer compressed data to $14\times 14\times c$, with $c$ being the bottleneck's dimension selected from the set $\mathcal{C}=\{4,8,16,32\}$. Then, assuming $32$ bits to represent each scalar, the number of bits transmitted was $n_b=14\times 14\times c\times 32\times B$, with $B=16$ being the batch size. Once the data was uploaded, it was first decoded. Then, the inference model was selected out of three different pre-trained ones that were fine tuned\footnote{https://docs.pytorch.org/vision/main/models.html}: \texttt{Mobilenet-v3-small} ($0.12$~GFLOPs), \texttt{Resnet-50} ($8.18$~GFLOPs), and \texttt{vit\_b\_16} ($33.7$~GFLOPs). For the purpose of efficient inference, these models were compiled with Torch-TensorRT\footnote{https://docs.pytorch.org/TensorRT/}. Then, their performance, in terms of delay, was tested on an NVIDIA A30 GPU. 
Data was collected for inference on $5000$ batches. The average delay was $[6,10,32]$~ms, respectively for \texttt{Mobilenet-v3-small}, \texttt{Resnet-50}, and \texttt{vit\_b\_16}. The decoding delay (whose average is $~4$ ms) was also considered. For the purpose of solving problem $\mathcal{P}$, that data were used to characterize $L_{\text{comp,b,m}},$ using the 98th percentile to run the optimization. 
Once the optimization was performed, true delay realizations were used for a Monte Carlo simulation to compute~\eqref{go_eff}. The quality metric used for the goal effectiveness was the number of correctly classified samples within a batch.

\textbf{Requirements.} We have set $L_{\max}=50$ ms and $Q_{\min}=11$ samples for the goal effectiveness (cf. \eqref{go_eff}). For the ISAC service, we have set $\sqrt{\text{CRB}_{\theta}^{\text{th}}}=0.3$ degrees and $R_{\text{dl}}^{\text{th}}=200$ Mbps. The parameter $\sigma$ in $\mathcal{P}$ was explored in $[0,1]$ with $1000$ linearly spaced values.

\textbf{Benchmark.} 
The core idea of our contribution is a compute-aware optimization, with radio resources and source coding optimized in conjunction with the knowledge of the ML model used for inference. We argue that this computation/model-awareness helps achieving a better trade-off between DL communication power and goal effectiveness under ISAC requirements. Therefore, as a benchmark, we propose a compute-unaware strategy, which assumes \texttt{vit\_b\_16} when optimizing, with the goal of ensuring the delay part of the goal effectiveness metric.

\textbf{Results discussion.} Figure~\ref{fig:tradeoff_1} illustrates the goal effectiveness as a function of the DL transmit power, obtained by varying the weighting parameter $\sigma$ in $\mathcal{P}$. The different colors are related to the used inference model, whose computational complexity in terms of GFLOPs appear in the legend. Solid lines represent our computation-aware approach, while the dashed dotted lines indicate the disjoint benchmark. First, the strong relation between physical-layer parameters of the ISAC service and application performance for EI can be noticed. In particular, the goal effectiveness is a non-decreasing function of the DL transmit power. This is due to the fact that, a higher goal effectiveness requires more time resources for the EI service, thus constraining the DL ISAC service to increase its transmit power due to constraints $(a)$ and $(b)$. Also, a better trade-off can be obtained if more computing resources are used (up to $33.7$~GLOPs). Of course, this typically incurs higher cost, e.g., in terms of energy consumption. Further, we notice the superiority of our computation-aware optimization against the disjoint benchmark. For example, for $8.18$~GFLOPs, higher goal effectiveness can be obtained with almost half of the DL transmit power. In this case, the same performance is achieved with \texttt{vit\_b\_16}, as it is used as information for the computation-unaware optimization. In Fig.~\ref{fig:tradeoff_2}, we illustrate the same trade-off obtained with our strategy, with two different CBR requirements, as shown in the legend. This figure showcases how relaxing the sensing requirements strongly affects the trade-off between DL transmit power and EI-basedd goal effectiveness. Also importantly, it is observed that, for DL transmit power above $80$~mW, both \texttt{Resnet-50} and \texttt{vit\_b\_16} achieve goal effectiveness above $80\%$, therefore, a fraction of compute power can be only used, provided that the respective lightweight ML model is available at the ES.

\section{Conclusions}
This paper elaborates on the complex interaction between ISAC and edge intelligence services, when they share wireless network resources. After showing how metrics of these coexistent services are coupled, we formulated a joint optimization problem for them via cross-layer parameters. It was showcased that the proposed joint, compute-aware approach helps achieving a better trade-off between DL user rate and, therefore, the feasible angular resolution for sensing and goal effectiveness for EI. Beyond this preliminary study, several research directions emerge: \textit{i}) dynamic optimization adapting to contextual factors, such as target behavior, and enabling online inference model selection; and \textit{ii}) investigation of synergistic coexistence, e.g., leveraging sensing data to enhance EI performance rather than treating sensing and computation as competing services.

\bibliographystyle{IEEEtran}
\bibliography{Main}

\end{document}